\documentclass[twocolumn,showpacs]{revtex4}
\usepackage{graphicx}
\usepackage{bm}
\usepackage{amssymb,amsmath}
\usepackage[usenames]{color}
\usepackage{epsfig}
\usepackage{hyperref}
\usepackage{subfigure}
\usepackage[sans]{dsfont}
\hypersetup{colorlinks=true, citecolor=blue,
linkcolor=blue,urlcolor=blue }

\begin{document}

\title{Anomalies in conductance and localization length of disordered ladders}
\author{Reza Sepehrinia}
\affiliation{Department of Physics, University of Tehran, Tehran
14395-547, Iran} \email{sepehrinia@ut.ac.ir}

\affiliation{School of Physics, Institute for Research in
Fundamental Sciences, IPM, 19395-5531 Tehran, Iran}

\begin{abstract}

We discuss the conditions under which an anomaly occurs in
conductance and localization length of Anderson model on a lattice.
Using the ladder hamiltonian and analytical calculation of average
conductance we find the set of resonance conditions which
complements the $\pi$-coupling rule for anomalies. We identify those
anomalies that might vanish due to the symmetry of the lattice or
the distribution of the disorder. In terms of the dispersion
relation it is known from strictly one-dimensional model that the
lowest order (i.e., the most strong) anomalies satisfy the equation
$E(k)=E(3k)$. We show that the anomalies of the generalized model
studied here are also the solutions of the same equation with
modified dispersion relation.

\end{abstract}
 \maketitle

\section{Introduction}

Several studies have been done on the anomalous behavior of
one-dimensional models of Anderson localization. The anomaly is
mainly understood as a resonance effect that occurs upon
commensurability of particle wavelength with lattice spacing of the
background periodic potential.

The main tool to investigate the anomalous behavior has been the
weak disorder expansion of Lyapunov exponent \cite{Derrida}. Anomaly
is signaled by divergence of certain orders in the expansion. The
divergences can be overcome by taking into account the degeneracies
and implementing the correct perturbation theory \cite{Kappus}. One
then finds corrections of sub-leading order and the largest
correction is obtained at the center of the band. At the band edge
of pure system however all orders of expansion diverge as it
requires a nonanalytic dependence on disorder strength
\cite{Derrida}.

The problem of anomaly has attracted considerable interest within
the mathematical community. There exist several rigorous results
regarding the existence and classification of the anomalies
\cite{Bovier, Campanino, Schulz}. Besides being mathematically
subtle, anomaly is accompanied with interesting physical situations.
The coupled statistical evolution of the phase and amplitude of
waves along the system implies a connection between the anomaly and
violation of the random phase hypothesis \cite{Lambert}. Namely, one
obtains the result of non-degenerate perturbation theory by assuming
uniform phase distribution \cite{Douglas}. Another property that
follows from random phase assumption is log-normal distribution of
conductance \cite{Anderson} which does not hold at the anomaly
\cite{Schomerus}. This has been considered as violation of single
parameter scaling theory \cite{AALR} which is one of the fundamental
frameworks of localization. Moreover the violation of reflection
phase randomization at the anomaly which implies a phase relation
between incident and reflected waves suggests applications in
designing photonic or electronic filters \cite{Titov}.

Perturbative calculation of Lyapunov exponents has been extended to
include multiple chains \cite{Hermann} and next-nearest-neighbor
hopping terms \cite{Reza}. Similarly the breakdown of perturbation
theory determines some exceptional energies. However the condition
for resonance is not always a simple commensurability in such cases.
In contrast to the strictly one dimensional model, commensurability
of a combination of wave vectors corresponding to different
transmission channels can result in the anomaly
\cite{Hermann,Reza,Hong}.

Then the natural question that arises is whether or not there exist
a unified way of describing the anomalies in terms of fundamental
properties of the system. There have been efforts to ascribe the
anomaly to the symmetries of the hamiltonian but this approach has
remained limited to simple models \cite{Deych}. Recently a
diagrammatic explanation of anomalous behavior is provided based on
scattering theory methods \cite{Luca}. It turns out to be a useful
method in application to more complicated cases. Accordingly, the
anomaly is the result of coherent interference of scattering
amplitudes from different lattice points. To have such coherency,
specific relation between the wavevectors of left-going and
right-going waves propagating in the chain is needed. It is shown
that center of the band anomaly requires $k^{+}-k^{-}=\pm\pi$, where
$k^{\pm}$ are the wavevectors of the right and left going waves.
This also generalizes to the case of multiple coupled chains with
several energy bands. The so called $\pi$-coupling,
$k_{\mu}^{+}-k_{\nu}^{-}=q\pi$ with integer $q$, between different
bands also results in the similar anomalies. This result shows that
even though the anomalies do not exist in the density of states of
pure system, the structure of the energy bands tells us where they
would appear by turning on the disorder.

The above mentioned rule for resonance is obtained for the special
type of tight-binding models for which the hamiltonian of the unit
cell commutes with the hopping matrix (see Eq. (\ref{NNN})). As a
result the eigenfunctions of corresponding pure hamiltonians are
plane waves, $\Phi_{\nu}(n)= e^{ink}\bm{\chi}_{\nu}$, with
$k$-independent amplitudes $\bm{\chi}_{\nu}$. Here $\bm{\chi}_{\nu}$
is a vector with the dimension of number of atoms in the unit cell
and $\nu$ represents the energy bands.

In this paper we discuss a model which is less symmetric and does
not possess the above property. We show that more complete forms of
anomalous couplings can be realized in this model. These couplings
are discussed in the section \ref{sec-cond}. In the section
\ref{sec-loc} the results are compared with those of localization
length. We show that anomalous wavevectors are roots of the same
equation which was obtained for 1D Anderson model. In the section
\ref{sec-ladd} we examine the possibility of resonances due to new
couplings in the symmetric ladder model. We show that new couplings
can be seen even in this model by introducing inhomogeneity in the
disorder.

\section{Model and unperturbed Green's Function}
\begin{figure}[t]
\epsfxsize8truecm \epsffile{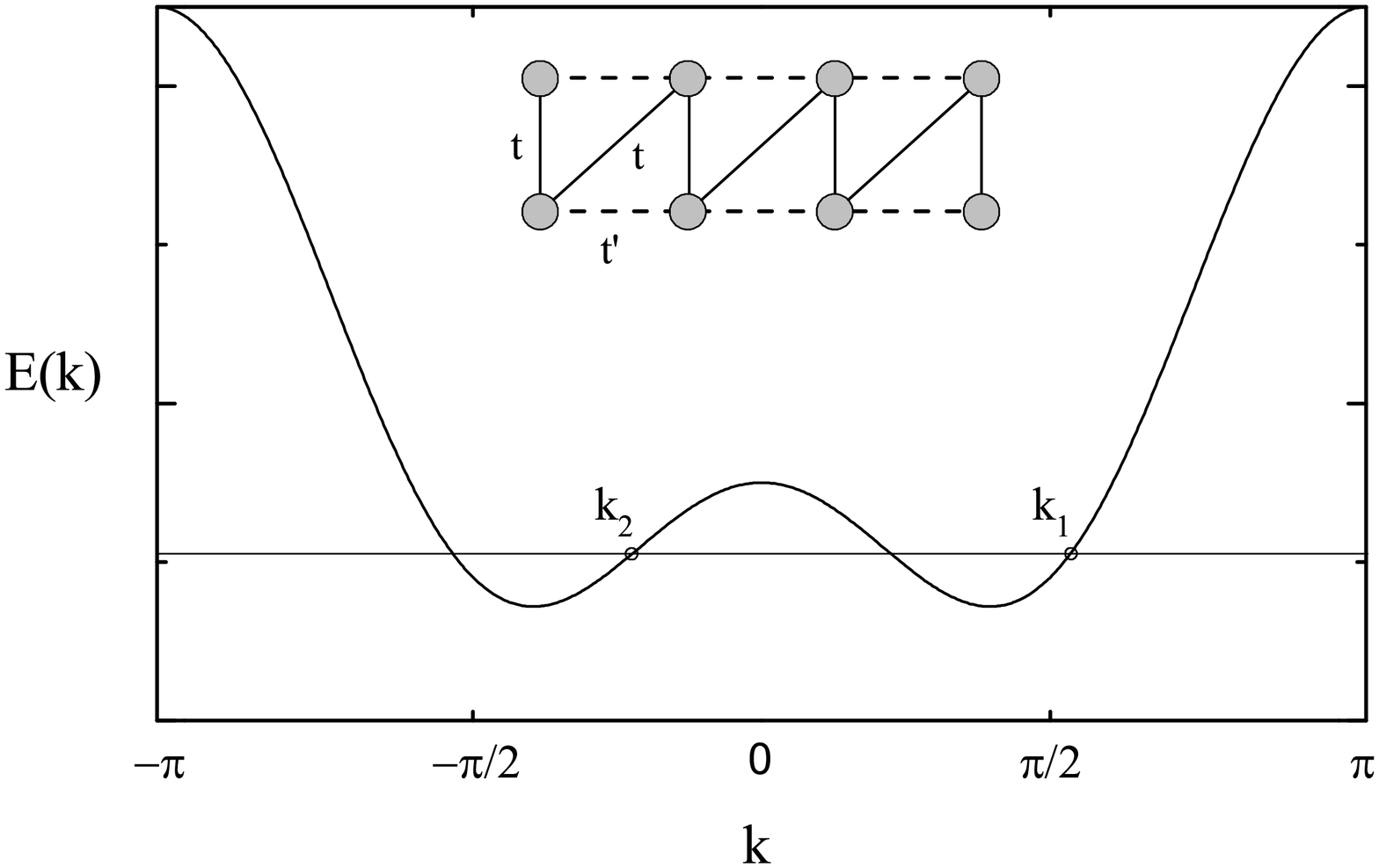} \caption{Dispersion relation
(\ref{dispersion}) for $|\frac{t}{t'}|<4$ and the lattice. $k_1$ and
$k_2$ are conjugate wave vectors carrying equal energy in the
positive direction.}\label{disp-latt}
\end{figure}
We consider the Anderson model with next-nearest-neighbor hopping
which can be viewed as a double chain (see Fig. \ref{disp-latt})
with nearest-neighbor hopping
\begin{equation}\label{Sch}
    T \Phi(n+1)+U \Phi(n) + T^{\dag}\Phi(n-1)=E\Phi(n), \nonumber
\end{equation}
\begin{equation}\label{NNN}
    U=\left(
\begin{array}{cc}
    0 & t \\
    t & 0
\end{array}\right)
, \ \ \ T=\left(
\begin{array}{cc}
    t' & t \\
    0 & t'
\end{array}\right),
\end{equation}
where $t$ and $t'$ are nearest-neighbor and next-nearest-neighbor
hopping integrals, respectively. As we mentioned $[U,T]\neq0$ and
the amplitudes of eigenfunctions are $k$-dependent
\begin{equation}\label{amp}
    \bm{\chi}_1=\frac{1}{\sqrt{2}}\left(
\begin{array}{c}
    1 \\
    e^{i\frac{k}{2}}
\end{array}\right),  \bm{\chi}_2=\frac{1}{\sqrt{2}}\left(
\begin{array}{c}
    1  \\
    -e^{i\frac{k}{2}}
\end{array}\right).
\end{equation}
It is clear from (\ref{amp}) that these eigenfunctions are also
plane wave solutions for single chain with halved lattice constant.

To avoid the matrix notation we use the single chain picture. In the
real space basis the hamiltonian corresponding to Eq. (\ref{NNN})
has the following matrix elements
\begin{eqnarray}\label{hamiltonian}
   H_{0_{nm}}=t(\delta_{n,m+1}+\delta_{n,m-1})+t'(\delta_{n,m+2}+\delta_{n,m-2}),
\end{eqnarray}
where $\delta_{ij}$ is the Kronecker delta. Eigenfunctions of this
hamiltonian are $\phi^{\nu}(n)=e^{ik_{\nu}n}$ and its eigenvalues
satisfy the dispersion relation
\begin{equation}\label{dispersion}
    E(k)=2t\cos k+2t'\cos 2k, \ \  -\pi\leq k\leq\pi.
\end{equation}
The Green's function of the model (\ref{NNN}) and similar models
have been already discussed in the literature
\cite{Bahurmuz,Schwalm}. Appropriate expression for real space
matrix elements of Green's function $G_0^+=(E-H_0+i\eta)^{-1}$ can
be obtained using the representation in the basis of eigenfunctions
of $H_0$,
\begin{equation}\label{Integral}
    G_{0mn}^+=-\frac{i}{2\pi}\oint\frac{z^{|m-n|-1}dz}{E-E(k(z))+i\eta},
\end{equation}
where the integral is taken along the unit circle $z=e^{ik}$. With
the dispersion (\ref{dispersion}) the integrand has two pairs of
poles $e^{\pm ik_1}$ and $e^{\pm ik_2}$. Each pair corresponds to
either an open or a closed channel, depending on if they lie on the
unit circle or not. One pole from each pair which lies inside the
circle contributes to the integral (\ref{Integral}).

\section{Transmission and Conductance}\label{sec-cond}
We use the scattering approach based on Lippmann-Schwinger equation.
As usual we consider an infinite chain with a disordered segment of
size $L$ in the middle. Motion of particle is governed by full
hamiltonian $H=H_0+V$ in the disordered part and the pure
hamiltonian $H_0$ in two perfect leads. We will discuss the weak
disorder limit so the random part of hamiltonian, $V$, will be
assumed as perturbation. Starting with an incident wave $\Phi$ from
the left in specific channel we end up with the scattering state
$\Psi$ satisfying the Lippmann-Schwinger equation
$\Psi=\Phi+G^{+}_{0}V\Psi$. It can be rewritten as
\begin{equation}\label{LS}
    \Psi=(1-G^{+}_{0}V)^{-1}\Phi.
\end{equation}
Away from the scattering center this state will have the asymptotic
form $\phi^{\mu}(n)  + \sum^{-}_{\nu}r_{\mu\nu} \phi^{\nu}(n)$ in
the left lead and $\sum^{+}_{\nu} t_{\mu\nu} \phi^{\nu}(n)$ in the
right lead. The $+$ and $-$ signs denote the restriction of
summations to right and left-going waves, respectively. At distances
which are not far from the scattering region the evanescent modes
should also be included in the later sum
\cite{Bagwell,Heinrichs-ev,Gasparian}. The conductance will be
obtained using the Landauer formula upon calculation of reflection
$r_{\mu\nu}$ and transmission $t_{\mu\nu}$ amplitudes (not to be
confused with hopping integrals $t,t'$).

For diagonal disorder, which will be considered here, the matrix
elements of the perturbation are
\begin{equation}\label{Vnm}
V_{nm}=\Big\{\begin{array}{c}
  w \varepsilon_n \delta_{nm}, \ \ \ \  1\leq n,m \leq L, \\
  0 \ \ \ \ \text{otherwise},
\end{array}
\end{equation}
where $w$ is the disorder strength and $\varepsilon_n$s are
uncorrelated random variables with symmetric distribution (for
simplicity) and finite moments. Up to fourth order of disorder
strength from Eq. (\ref{LS}) we have
\begin{eqnarray}\label{psi}
    \psi^{i}(r)&=& e^{irk_i}  +  w \sum_{n=1}^{L}
G_{0rn}\varepsilon_n e^{ink_i}    \nonumber \\ &&+ w^2
\sum_{m,n=1}^{L}G_{0rm}G_{0mn} \varepsilon_m \varepsilon_n e^{ink_i}
\nonumber \\ &&+ w^3
\sum_{l,m,n=1}^{L}G_{0rl}G_{0lm}G_{0mn}\varepsilon_l\varepsilon_m
\varepsilon_n e^{ink_i}         \nonumber \\ &&+ w^4
\sum_{f,l,m,n=1}^{L}G_{0rf}G_{0fl}G_{0lm}G_{0mn}\varepsilon_f\varepsilon_l\varepsilon_m\varepsilon_n
e^{ink_i}, \nonumber \\
\end{eqnarray}
where $r=L+1,L+2$.

\subsection{Two open channels}
\label{twoch}

For hopping ratios $|\frac{t}{t'}|<4$, there are four Fermi points
at a given energy in the interval $-2t'-\frac{t^2}{4t'}<E<2t'-2t$
which two of them have positive group velocity (i.e., positive slope
in Fig. \ref{disp-latt}) and contribute to the Green's function
(\ref{Integral})
\begin{equation}\label{Green2}
    G_{0mn}^+=-i\Big(\frac{e^{ik_1|m-n|}}{v_1}+\frac{e^{ik_2|m-n|}}{v_2}\Big),
\end{equation}
where $k_i$s are real roots of $E-E(k)=0$ with positive group
velocities $v_i=\frac{\partial E}{\partial k}|_{k=k_i}$.

Suppose now the incident wave from left is $\phi^i$. The scattering
state would then have the form $\psi^i(n)=t_{i1}
e^{ik_1n}+t_{i2}e^{ik_2n}$ in the right lead. To obtain two
amplitudes $t_{i1},t_{i2}$ in this expression we need to know
$\psi^i$ at two lattice points say $\psi^i(L+1)$ and $\psi^i(L+2)$,
\begin{equation}\label{FT}
    \left(\begin{array}{c}
                   t_{i1} \\
                   t_{i2} \\
                 \end{array}
               \right)
=\left(
                  \begin{array}{cc}
                    e^{ik_1(L+1)} & e^{ik_2(L+1)} \\
                    e^{ik_2(L+2)} & e^{ik_2(L+2)} \\
                  \end{array}
                \right)^{-1}\left(
                              \begin{array}{c}
                                \psi^i(L+1) \\
                                \psi^i(L+2) \\
                              \end{array}
                            \right).
\end{equation}
From (\ref{FT}), (\ref{psi}) we get
\begin{eqnarray}\label{tij}
    t_{ij}&=& \delta_{ij}  -i  \frac{w}{v_j} \sum_{n=1}^{L}
\varepsilon_n e^{in(k_i-k_j)}    \nonumber \\ &&-i \frac{w^2}{v_j}
\sum_{m,n=1}^{L}G^+_{0mn} \varepsilon_m \varepsilon_n
e^{i(nk_i-mk_j)} \nonumber \\ &&-i \frac{w^3}{v_j}
\sum_{l,m,n=1}^{L}G^+_{0lm}G^+_{0mn}\varepsilon_l\varepsilon_m
\varepsilon_n e^{i(nk_i-lk_j)}         \nonumber \\ &&-i
\frac{w^4}{v_j}
\sum_{f,l,m,n=1}^{L}G^+_{0fl}G^+_{0lm}G^+_{0mn}\varepsilon_f\varepsilon_l\varepsilon_m\varepsilon_n
e^{i(nk_i-fk_j)}. \nonumber \\
\end{eqnarray}
Conductance is given now in terms of transmission amplitudes
\begin{equation}\label{cond}
    g=\frac{2e^2}{h}\sum_{i,j=1}^{2}\rho_iv_j|t_{ij}|^2,
\end{equation}
where $\rho_i=\frac{1}{2\pi}|\frac{\partial E}{\partial
k}|^{-1}_{k=k_i}$ is density of states. Four ensemble averaged
transmission probabilities up to fourth order of disorder strength
are obtained as follows
\begin{widetext}
\begin{eqnarray}\label{t11-2}
\langle|t_{11}|^2\rangle&=&1-\frac{\langle\varepsilon^2\rangle}{v_1^2}(1+2v)
Lw^2
+\frac{1}{2v_1^4}(1+v)^2(1+2v)\langle\varepsilon^4\rangle
Lw^4+\frac{1}{v_1^4}(3v^2+2v+1)\langle\varepsilon^2\rangle^2
L(L-1)w^4
\nonumber\\&&+\frac{2\langle\varepsilon^2\rangle^2}{v_1^4}\Big(v^3C(1,k_1-3k_2)+v^3C(1,k_1+3k_2)
+3v^2C(1,2k_1+2k_2)+v
C(1,3k_1-k_2)\nonumber\\&&+3vC(1,3k_1+k_2)+C(1,4k_1)
+(2v^3+3v^2)C(1,2k_2)+(v^3+v)C(1,k_1-k_2)\nonumber\\&&+(v^3+4v^2
+3v)C(1,k_1+k_2)+(2v^2+2v+1)C(1,2k_1)\Big)w^4,
\\
\langle|t_{22}|^2\rangle&=&\langle|t_{11}|^2\rangle|_{1\rightarrow
2, 2\rightarrow 1},
\label{t22-2}\\
\langle|t_{12}|^2\rangle&=&\frac{\langle\varepsilon^2\rangle}{v_2^2}
Lw^2
-\frac{1}{v_1^2v_2^2}(1+v)^2\langle\varepsilon^4\rangle
Lw^4-\frac{2v}{ v_1^2v_2^2}\langle\varepsilon^2\rangle^2L(L-1)w^4
-\frac{2\langle\varepsilon^2\rangle^2}{v_1^2
v_2^2}\Big(v^2C(1,k_1-3k_2) \nonumber\\&&-2vC(1,k_1-k_2)
+C(1,3k_1-k_2)+(v^2+2v)C(1,2k_2) +
(2v+1)C(1,2k_1)\nonumber\\&&+(v+1)^2C(1,k_1+k_2)\Big)w^4,
\label{t12-2}\\
\langle|t_{21}|^2\rangle&=&\langle|t_{12}|^2\rangle|_{1\rightarrow
2, 2\rightarrow 1},\label{t21-2}
\end{eqnarray}
\end{widetext}
where $v=\frac{v_1}{v_2}$ and $C(a,\theta)=\sum_{n<m}^L
a^{m-n}\cos[(m-n)\theta]$. Summations in (\ref{t11-2}-\ref{t21-2})
with $a=1$ can be written in the following closed form
\begin{eqnarray}\label{sumc}
    C(1,\theta)&=&\sum_{n<m}^L \cos[(m-n)\theta]\nonumber\\
               &=&\frac{1-\cos L\theta}{2(1-\cos \theta)}-\frac{1}{2}L.
\end{eqnarray}
For $\theta\neq 2n\pi$ the second term dominates and the sum behaves
linearly with size at $L\rightarrow\infty$ but this is not the case
if $\theta=2n\pi$ for which the dependence on $L$ is quadratic.

At the limit $L\rightarrow\infty$ the applicability of perturbation
theory is determined by the leading terms which are proportional to
$w^2L$ and $(w^2L)^2$ in the second and fourth orders, respectively.
It can be easily seen that the leading term in higher orders is
proportional to $(w^2L)^{\frac{n}{2}}$  where $n$ is order of
perturbation. We need to decrease the disorder strength $w$ as we
are increasing the $L$, such that
\begin{equation}\label{condition}
   \frac{w^2L\langle\varepsilon^2\rangle}{\text{min}\{v_1^2,v_2^2\}}\ll
   1.
\end{equation}
Consequently the terms proportional to $w^4L$ in the fourth order
will vanish in such limit. Therefor the summations in each
transmission probability will give a finite contribution to the
fourth order term only if $\theta=2n\pi$, with $n$ being an integer.
This leads a narrow peak at some special energies which is called
anomaly.

According to dispersion relation (\ref{dispersion}) and depending on
the ratio $|\frac{t}{t'}|$, combinations which can satisfy the
condition $\theta=2n\pi$ are

\begin{eqnarray}\label{theta}
\theta=4k_1,2k_2&,&4k_2,k_1+k_2,2(k_1+k_2)\nonumber
\\&&,3k_1+k_2,k_1+3k_2,3k_1-k_2,
\end{eqnarray}
where
$k_{1,2}=\mp\arccos\left[\frac{1}{4}(-\frac{t}{t'}\pm\sqrt{4\frac{E}{t'}+(\frac{t}{t'})^2+8})\right]$.
For $|\frac{t}{t'}|<\frac{4\sqrt{6}}{9}$  those terms which lead
resonance inside the interval $-2t'-\frac{t^2}{4t'}<E<2t'-2t$ are
$\theta=4k_1,3k_1+k_2,k_1+3k_2,3k_1-k_2$. The rest of them get
resonant at the edges $E=-2t'-\frac{t^2}{4t'}$ or $E=2t'-2t$.

By inserting the transmission probabilities in conductance (Eq.
(\ref{cond})), we see that the terms containing $C(1,3k_1-k_2)$
cancel each other. So one of the above four terms vanishes and there
will be three peaks in conductance with the relative heights of
$\frac{A_{3k_1+k_2}}{A_{4k_1}}=4v,
\frac{A_{k_1+3k_2}}{A_{4k_1}}=4v^3$.

\subsection{One open and one closed channel} \label{onech}

In the rest of the energy band $2t'-2t<E<2t'+2t$ there is one pair
of fermi points when $|\frac{t}{t'}|<4$. One of the poles that
contribute to the Green's function integral is on the real axis
which corresponds to $k=i\kappa$ (positive side) or $k=i\kappa+\pi$
(negative side) and the other one is on the unit circle. The former
has real contribution to the Green's function
\begin{equation}\label{Green1}
    G_{0mn}^+=-i\frac{e^{ik|m-n|}}{v}+(\pm1)^{|m-n|}\frac{e^{-\kappa|m-n|}}{u},
\end{equation}
where $\kappa>0$ since the pole is inside the unit circle and
$u=-z\frac{\partial E}{\partial z}|_{z=\pm e^{-\kappa}}$.

In this case the wave function in the right lead is a combination of
a propagating and an evanescent mode. Transmission occurs through
one channel with ensemble averaged probability
\begin{widetext}
\begin{eqnarray}\label{t11-1}
\langle|t_{11}|^2\rangle&=&1-\frac{\langle\varepsilon^2\rangle}{v_1^2}
Lw^2
+\Big(\frac{1}{v_1^4}-\frac{3}{v_1^2
u^2}\Big)\langle\varepsilon^4\rangle Lw^4
+\Big(\frac{2}{v_1^4}L(L-1)
-\frac{8}{v_1^2u^2}C(-e^{-\kappa},k_1)-\frac{4}{v_1^2u^2}C(e^{-2\kappa},0)\nonumber\\&&-\frac{4}{v_1^2u^2}C(e^{-2\kappa},2k_1)
+\frac{2}{v_1^2}(\frac{1}{v_1^2}-\frac{1}{u^2})C(1,2k_1)+\frac{2}{v_1^4}C(1,4k_1)-\frac{4}{v_1^3u}S(1,2k_1)
-\frac{4}{v_1^3u}S(-e^{-\kappa},k_1)\nonumber\\&&-\frac{4}{v_1^3u}S(-e^{-\kappa},3k_1)\Big)\langle\varepsilon^2\rangle^2w^4,
\end{eqnarray}
\end{widetext}
where $S(a,\theta)=\sum_{n<m}^L a^{m-n}\sin[(m-n)\theta]$. The
asymptotic behavior of summations like $S(1,\theta)$, $S(\pm
e^{-\kappa},\theta)$ and $C(\pm e^{-\kappa},\theta)$ should be
determined in order to obtain the limiting value of (\ref{t11-1})
for long chain. Unlike the sum of cosines, $S(1,\theta)$ can not
give a quadratic dependence on size
\begin{eqnarray}\label{sums}
    S(1,\theta)&=&\sum_{n<m}^L \sin[(m-n)\theta]\nonumber\\
               &=&-\frac{\sin L\theta}{2(1-\cos \theta)}+\frac{\sin \theta}{2(1-\cos \theta)}L.
\end{eqnarray}
Other sums with $a=\pm e^{-\kappa}$ and $\kappa>0$ at most will have
the following value
\begin{eqnarray}\label{sumkappa}
    S(\pm e^{-\kappa},\theta),C(\pm e^{-\kappa},\theta)   \leq   \sum_{n<m}^L (e^{-\kappa})^{m-n}\nonumber\\
    \underset{L \to \infty}\rightarrow -\frac{1}{1-e^{-\kappa}}L,
\end{eqnarray}
Therefor the only possibility to get $L^2$, again comes from
$C(1,2n\pi)$, otherwise we will have linear or oscillatory
asymptotic behavior. Moreover the only term which satisfy this
condition is $C(1,4k_1)$ with $4k_1=-2\pi$ at $E=-2t'$ which results
in the enhancement of transmission with the amount of
$\frac{1}{v_1^4}\langle\varepsilon^2\rangle^2 (Lw^2)^2$.

\subsection{Multi-channel case} \label{}

The generalization of the model (\ref{hamiltonian}) for the next
nearest neighbors is also straight forward only by replacing the
green's function with
\begin{equation}\label{}
    G_{0mn}^+=-i\sum_{\nu}\frac{e^{ik_{\nu}|m-n|}}{v_{\nu}}.
\end{equation}
Then different resonant wave vector combinations are expected to be
found.

\section{Localization length}\label{sec-loc}

In earlier publication [\onlinecite{Reza}] we have discussed the
perturbative calculation of the localization length for the model
(\ref{NNN}). There we assumed perturbative solutions
$\frac{\Psi_{n+1}}{\Psi_n}= e^{ik}e^{B_n w + C_n w^2 + \cdots}$ with
the growth rate
\begin{equation}\label{LE}
    \gamma(E)=\frac{1}{\xi}=w \langle
B\rangle+w^2\langle C\rangle+\cdots.
\end{equation}
We showed that the correlation function $\langle B_n B_m\rangle$ has
poles corresponding to anomalous energies as well as band edges of
pure hamiltonian. As an example in the interval
$|\frac{t}{t'}|<\frac{4\sqrt{6}}{9}$ there are four poles that
correspond to anomalous energies on the real $k$ axis
\begin{equation}\label{}
k=\frac{\pi}{2}
,\ \arccos\Big(\sqrt{\frac{2}{3}}\cos\left[\frac{u}{3}+n\frac{\pi}{3}\right]\Big), \ n=0,1,2, \\
\end{equation}
where $\cos u=-t/[2(\frac{2}{3})^{\frac{3}{2}}t']$ and $0<u<\pi$. In
relation to the resonance conditions that was obtained in the
previous section these wave vectors satisfy the equations $4k=2\pi,
3k+k'=0, 3k+k'=\pm 2\pi, 3k-k'=\pm 2\pi$, respectively, where $k'$
is the conjugate wave vector to $k$ (see Fig. \ref{Fig-loclen}). The
last case was absent in the numerical results for localization
length \cite{Reza}. It is the one that we showed in previous section
appears in partial interchannel transmission but different
contributions cancel each other in the conductance.

The above relations between $k$ and $k'$ together with $E(k)=E(k')$
lead the equation
\begin{equation}\label{}
    E(k)=E(3k),
\end{equation}
for the poles. This equation was first obtained in 1D Anderson model
with only nearest neighbor hopping \cite{Kappus}.

\section{symmetric ladder model and inhomogeneous
disorder}\label{sec-ladd}

The ladder model with symmetric unit cell hamiltonian and diagonal
hopping matrix
\begin{equation}\label{ladder}
U=\left(
\begin{array}{cc}
    0 & t \\
    t & 0
\end{array}\right),
\ \ T=\left(
\begin{array}{cc}
    t' & 0 \\
    0 & t'
\end{array}\right),
\end{equation}
exhibits anomalies resulted from intra-band and inter-band
$\pi$-coupling \cite{Luca}. In a later numerical study \cite{Nguyen}
it is shown that extra anomalies appear by taking different widths
of disorder in two chains. These new anomalies were also attributed
to the $\pi$-coupling of bands but at two different energies. We
show that they can be described only by the coupling of waves in a
single energy that we obtained in the section \ref{sec-cond}.

We consider the following random potential which is studied
numerically in the reference \cite{Nguyen}
\begin{equation}\label{pot-ladder}
    \hat{V}_{n}=\left(
\begin{array}{cc}
    w\varepsilon_n & 0 \\
    0 & w\eta_n
\end{array}\right),
\end{equation}
where $n$ indicates a column with two atoms. Eigenfunctions of
unperturbed hamiltonian are
\begin{eqnarray}
    \Phi_1(n)=e^{ik_1n}\bm{\chi}_1, \ \ \Phi_2(n)=e^{ik_2n}\bm{\chi}_2,
\end{eqnarray}
each of which corresponds to a transmitting channel and $
    \bm{\chi}_1=\frac{1}{\sqrt{2}}(
\begin{smallmatrix}
    1 \\
    1
\end{smallmatrix}),  \bm{\chi}_2=\frac{1}{\sqrt{2}}(
\begin{smallmatrix}
    1  \\
    -1
\end{smallmatrix})$.
In this basis the green's function is given by
\begin{equation}\label{Green-ladder}
    \hat{G}^+_{0nm}=-i\frac{e^{ik_1|m-n|}}{v_1}\bm{\chi}_1^{\dag}\bm{\chi}_1-i\frac{e^{ik_2|m-n|}}{v_2}\bm{\chi}_2^{\dag}\bm{\chi}_2.
\end{equation}
Transmission coefficients can be obtained in the similar way that we
did in the section \ref{sec-cond} by generalizing Eq. (\ref{psi}) to
the matrix form. We do not give the full expressions of them and
only look for the missing resonant terms arising from
$\theta=3k_1\pm k_2, k_1+3k_2$. Such terms come from a fourth order
term like (say in $|t_{11}|^2$)
\begin{eqnarray}\label{}
    \langle1|\hat{G}^{+}_{0rm}\hat{V}_m
    \hat{G}^{+}_{0mn}\hat{V}_n\hat{G}^{+}_{0nm}\hat{V}_m\hat{G}^{+}_{0mn}\hat{V}_n|1\rangle+c.c.
\end{eqnarray}
from which we can get a term proportional to
\begin{eqnarray}\label{}
      (\varepsilon^2_n-\eta^2_n)(\varepsilon^2_m-\eta^2_m)\cos\left[(3k_1+k_2)|m-n|\right],
\end{eqnarray}
and after ensemble averaging it is proportional to
$(\langle\varepsilon^2\rangle-\langle\eta^2\rangle)^2$ which will
disappear if we have
$\langle\varepsilon^2\rangle=\langle\eta^2\rangle$.
\begin{figure}[t]
\epsfxsize7truecm \epsffile{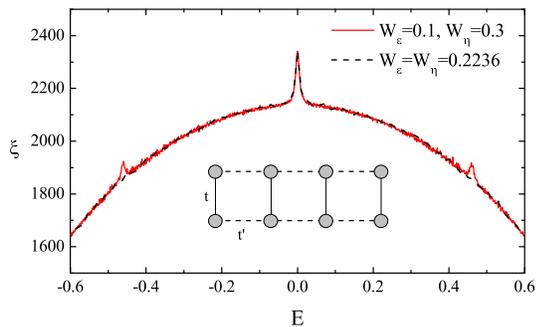} \caption{Numerical results of
localization length for symmetric ladder model ($t=t'=1$) for widths
of disorder on chains $W_{\varepsilon}=0.1, W_{\eta}=0.3$ (solid)
and $W_{\varepsilon}= W_{\eta}=0.2236$ (dashed). Note that the solid
curve has two small peaks at $E\approx\pm 0.46$ in addition to the
large peak at the band center.}\label{Fig-loclen}
\end{figure}

We have verified this result numerically by computing the
localization length (inverse of small Lyapunov exponent) of
symmetric ladder model using the transfer matrix method. Figure
\ref{Fig-loclen} shows the obtained results for localization length
of ladder with hopping integrals $t=t'=1$. Two different
configurations of disorder are considered, $W_{\varepsilon}=0.1,
W_{\eta}=0.3$ and $W_{\varepsilon}= W_{\eta}=0.2236$ where
$W_{\varepsilon},W_{\eta}$ are widths of uniformly distributed
potential on each chain. Both configurations are chosen to have
equal overall variance $(W_{\varepsilon}^2+W_{\eta}^2)/24$ and
consequently equal localization length up to second order of
perturbation. The difference in the localization lengths is of
fourth order which is apparent in the small peaks corresponding to
the couplings $3k_1+k_2=2\pi$,  $k_1+3k_2=2\pi$ in the case with
inhomogeneous disorder.

\section{conclusion}

We conclude that the $\pi$-coupling of energy bands is an instance
of wider forms of couplings leading the anomaly. Although these
couplings are necessary but not enough conditions for the appearance
of the resonances. Some couplings may not result in resonance due to
the following reasons (\textit{i}) symmetry of distribution of
values of random potential. As an example, the $k=\frac{\pi}{3}$
anomaly in strictly one-dimensional Anderson model that requires
asymmetric distribution of disorder \cite{Derrida} (\textit{ii})
symmetry of the lattice, such as in the symmetric ladder (Eq.
(\ref{ladder})) where the $k_1+3k_2$ coupling is absent but appears
in the asymmetric model (Eq. (\ref{NNN})) (\textit{iii}) spatial
symmetry of distribution of random potential, such as in the
symmetric ladder where $k_1+3k_2$ coupling shows up  by introducing
an inhomogeneous disorder.

\section{aknowledgement}
I would like to acknowledge the hospitality of ICTP where part of
this work was completed.

\end{document}